\documentclass[conference,a4paper]{APSIPA2025}
\usepackage{xcolor}
\usepackage{amsmath}
\usepackage{graphicx}
\usepackage[table]{xcolor}
\definecolor{lightblue}{RGB}{224,247,250}
\usepackage[table]{xcolor}  

\usepackage{array}    
\usepackage{booktabs}   
\usepackage{colortbl}   
\usepackage{xcolor}     
\usepackage{etoolbox}   

\newcolumntype{?}{!{\color{black}\vrule width 1.2pt}}

\makeatletter
\patchcmd{\toprule}   {\addlinespace} {\relax}{}{}
\patchcmd{\midrule}   {\addlinespace} {\relax}{}{}
\patchcmd{\bottomrule}{\addlinespace} {\relax}{}{}
\makeatother

\usepackage[hyphens]{url}   
\usepackage{xurl}            

\usepackage{colortbl, array, booktabs, xcolor}
\usepackage{multirow}
\usepackage{threeparttable}
\usepackage{booktabs}
\usepackage[backend=biber,style=ieee,]{biblatex}
\usepackage{footmisc}
\usepackage{geometry}
\usepackage{fancyhdr}

\fancypagestyle{firststyle}{
  \fancyhf{}
  \fancyhead[C]{2025 Asia Pacific Signal and Information Processing Association Annual Summit and Conference (APSIPA ASC)}
}

\begin{document}

\title{Segment Transformer: AI-Generated Music Detection via Music Structural Analysis}

\author{
\authorblockN{
Yumin Kim\authorrefmark{1} and
Seonghyeon Go\authorrefmark{1}
}

\authorblockA{
MIPPIA Inc. \\
\{kym, gsh\}@mippia.com}
}

\maketitle
\begingroup
  \renewcommand\thefootnote{\fnsymbol{footnote}}
  \setcounter{footnote}{1}
  \footnotetext{Equal contribution.}
\endgroup
\setcounter{footnote}{0}

\thispagestyle{firststyle}
\pagestyle{fancy}

\begin{abstract}

Audio and music generation systems have been remarkably developed in the music information retrieval (MIR) research field. The advancement of these technologies raises copyright concerns, as ownership and authorship of AI-generated music (AIGM) remain unclear. Also, it can be difficult to determine whether a piece was generated by AI or composed by humans clearly. To address these challenges, we aim to improve the accuracy of AIGM detection by analyzing the structural patterns of music segments. Specifically, to extract musical features from short audio clips, we integrated various pre-trained models, including self-supervised learning (SSL) models or an audio effect encoder, each within our suggested transformer-based framework. Furthermore, for long audio, we developed a segment transformer that divides music into segments and learns inter-segment relationships. We used the FakeMusicCaps and SONICS datasets, achieving high accuracy in both the short-audio and full-audio detection experiments. These findings suggest that integrating segment-level musical features into long-range temporal analysis can effectively enhance both the performance and robustness of AIGM detection systems.

\end{abstract}

\section{Introduction}
Recent advances in generative models have enabled high-quality content synthesis across diverse modalities, including images~\cite{yu2019attributing}, video~\cite{singer2022make}, audio~\cite{kreuk2022audiogen, liu2024audioldm}, and natural language~\cite{radford2018improving, li2024pre}. In particular, the emergence of intuitive and user-friendly tools such as Suno\footnote{\url{https://suno.com/}\label{fn:suno}} and Udio\footnote{\url{https://udio.com/}\label{fn:udio}} has made AI-based audio and music generation highly accessible, significantly lowering the barrier for the creation of AI-generated music (AIGM). This growing trend has underscored the urgent need for technical solutions capable of reliably distinguishing AIGM from human-composed music and other manipulated audio content, to prevent unethical use and protect music intellectual property~\cite{widder2024watching}. 

However, existing AI-generated audio detection~\cite{yi2023audio, liu2023asvspoof, yi2023add, masood2023deepfakes, patel2023deepfake} methods are often limited in local scope due to CNN architecture, making it hard to extract structural dependencies in full music. There's the SpecTTTra model suggested in~\cite{rahman2024sonics}, which successfully proposed a method to address this issue and extract long-range dependencies from the entire audio. However, it shows limited utilization of musical information such as BPM and downbeat for analyzing musical structure. 

To address these challenges, we propose a two-stage AIGM detection framework. In the stage-1, which detects AIGM from short audio segments, we begin by extracting meaningful representations from each segment. Specifically, we pair our Cross-Attention–based Transformer decoder (AudioCAT) with several self-supervised audio encoders~\cite{ma2023effectiveness, xie2021siamese, tak2022automatic, castellon2023codified} to implement a specialized detection approach for each feature extractor. We further integrate a pre-trained audio-effect encoder such as FXencoder~\cite{koo2023music} into our Transformer-based architecture (FXencoder-Segment).\\ 
\nobreak
\indent While various methodologies~\cite{mert} exist for extracting high-quality embeddings from short audio slices, the model’s strategy of directly mapping these slices to Transformer token embeddings still has potential for improvement. In stage 2, we slice full audio using beat tracking and apply our Segment Transformer model. It is designed to capture global song features through structural analysis. This approach improves the model's ability to understand relationships between musical features, enabling effective differentiation between the intentional compositional patterns of human musicians and the AI. Additionally, the segment-based approach maintains awareness of musical form while efficiently processing complete compositions, allowing the detection system to identify inconsistencies in musical structure development and motif progression throughout the entire music. \\
Our main contributions are as follows:\\
\textbf{(1) Segment Detection Architecture:} We propose AudioCAT, a cross-attention-based Transformer decoder architecture that employs feature extractors as encoders to derive meaningful representations from short audio segments. Additionally, we adopt pre-trained FXencoder, which encodes only audio-effect information, to demonstrate the potential of diverse feature extractors for effective AIGM detection.\\
\textbf{(2) Segment Transformer Architecture:} We propose a Segment Transformer model that employs beat-aware audio segmentation mechanisms to preserve temporal relationships and capture global song features through structural analysis of complete compositions.\\
\textbf{(3) Novel Stage-2 Framework:} We employ two segment detection models that extract audio features from short segments, while the Segment Transformer captures structural features from longer pieces. By combining these approaches for comprehensive musical analysis, we achieve superior detection performance and establish a new benchmark for full-audio AIGM detection.

\section{Methods}

\begin{figure}[t]           
  \centering
  \includegraphics[width=\columnwidth]{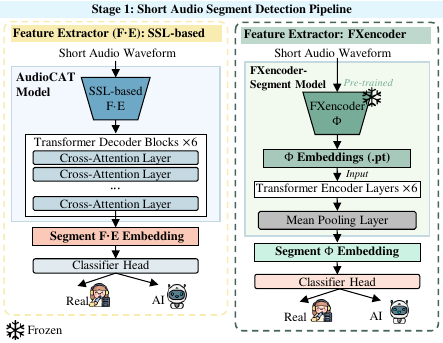}
  \caption{Overview of the proposed Stage-1: Short audio segment detection pipeline. We employ two types of feature extractors: self-supervised learning (SSL)–based models (Left) and an audio-effect–focused FXencoder~\cite{koo2023music}(Right).}
  \label{fig:segment_detection}
\end{figure}
\subsection{Stage 1: Short Audio Segment Detection Model}
\label{sec:segment-detection}

As shown in stage 1 of Figure \ref{fig:segment_detection}, we employ specialized feature extractors as the model’s encoder to detect short audio segments. Specifically, we use self-supervised learning (SSL) models and a pre-trained FXencoder~\cite{koo2023music} that captures audio-effect information. 
\subsubsection{AudioCAT} 
This architecture integrates an SSL-based feature extraction encoder with a~\textbf{C}ross-\textbf{A}ttention–based ~\textbf{T}ransformer decoder. The feature extraction encoder captures localized frequency patterns and textural elements from audio segments, producing spatially confined feature representations. Moreover, this encoder can be instantiated with a general-purpose SSL model trained on large-scale audio datasets or replaced by a task-specific encoder. In our architecture, the decoder latent states are used as queries,  while the encoder feature maps serve as keys and values, allowing the decoder to align its internal contextual representation with localized audio features. By leveraging the cross-attention mechanism to learn sequential representations, the model enhances its discriminative capacity, after which a classification head performs the AI-generated music (AIGM) detection task.

\noindent\textbf{Why Apply Cross-attention.} Cross-attention allows the decoder to strategically integrate features with its own internal representations, achieving better integration of local textural and global contextual information, where even short music segments contain crucial global patterns that are essential for effective AIGM detection performance.

\subsubsection{FXencoder-Segment Model}
We hypothesize that extracting audio-effect (FX) and music-specific features provides greater advantages for AIGM detection than relying on general audio representations. Consequently, we employ the FXencoder~\cite{koo2023music}—a specialized model originally designed for other audio tasks within our self-attention based Transformer encoder architecture. We use the pre-trained FXencoder to extract embedding vectors as input, keeping the feature extractor frozen during training so its representations remain fixed. While the cross-attention mechanism in AudioCAT attempts to “strategically combine” these features with decoder representations, this approach can be inefficient with a fixed FXencoder because the query remains. Therefore, a pure self-attention mechanism based on Transformer encoder layers is more appropriate, allowing all frozen features to directly interact and thereby effectively uncover relationships and patterns within the pre-trained embedding space.

\subsection{Feature Extractor for Segment Detection Model}
\label{sec:feature}

As shown in Figure~\ref{fig:segment_detection}, we selected both self-supervised learning (SSL) based pre-trained models and the FXencoder~\cite{koo2023music}, which we determined to provide essential characteristics for AIGM detection. 

The AudioCAT architecture is compatible with any feature extractor that produces meaningful embeddings from short audio segments. In our methods, we use Wav2vec~\cite{wav2vec}, Music2vec~\cite{limap}, and MERT~\cite{mert} as the feature extraction encoders. The feature extractor models used in our experiments are described as follows. 

\begin{figure*}[t!]
  \centering
  \includegraphics[width=\textwidth]{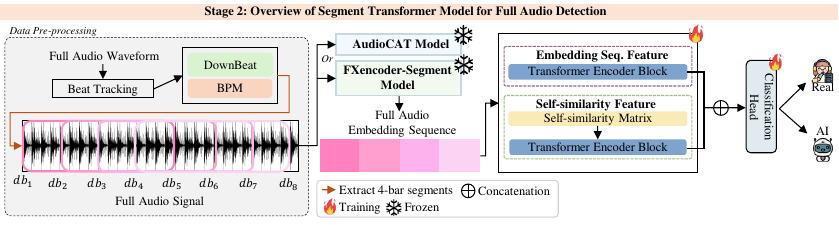}
  \caption{Overview of the proposed Stage-2: Segment Transformer for full audio segment detection pipeline. }
  \label{fig:segment_transformer}
\end{figure*}

\noindent\textbf{Wav2vec 2.0.} Existing spoofing detection~\cite{xie2021siamese, wang2021investigating} models primarily use features such as Mel-spectrogram or Mel-Frequency Cepstral Coefficient (MFCC)~\cite{tak2022automatic}, while Wav2vec-2.0~\cite{wav2vec, xu2021self} processes raw waveforms directly, allowing for the learning of more robust representations. Compared to conventional spoofing detection models~\cite{yi2023audio, liu2023asvspoof, yi2023add, masood2023deepfakes, patel2023deepfake}, it offers better generalization across diverse speech data. We employ a pre-trained \texttt{Wav2vec 2.0-Base} model that extract 512-dimensional SSL feature representations.

\noindent\textbf{Music2vec.} This pre-trained model~\cite{limap} adopts the pre-training paradigm of the speech version of data2vec-1.0~\cite{data2vec}. Specifically adapted for the music domain, this continuous prediction model employs a 1D Convolutional Embedder to provide waveform input to both the teacher and student Transformer models, enabling more effective representation learning for music data. We use a pre-trained \texttt{Music2Vec-v1(95M)} model with 95M parameters that extract 512-dimensional feature representations.

\noindent\textbf{MERT.} A large-scale SSL-based music understanding model~\cite{mert} is designed to address the challenges in modeling musical knowledge, particularly tonal characteristics and pitched attributes, which have hindered the extensive exploration of SSL applications in music audio. We use the \texttt{MERT-v1(95M)} model, which is designed to generalize better across more tasks and has been trained on a dataset to achieve superior performance compared to previous models that extracting 768-dimensional feature representations.

\noindent\textbf{FXencoder.} Music contains unique structural and harmonic patterns that fundamentally differ from general audio signals. The FXencoder is designed specifically to extract mixing and mastering features from audio files. Unlike SSL-based models that learn general audio patterns, the FXencoder focuses on uncovering production details and processing effects in music recordings, adding unique mixing and mastering-related features that are advantageous for AIGM detection. This capacity to extract production-related characteristics is valuable for analyzing how audio was produced and helps distinguish human-produced music from AI-generated compositions. FXencoder extracts a 2048-dimensional feature representation.

The two proposed frameworks are not limited to the specific models presented here; this study can accommodate any architecture capable of extracting meaningful embeddings from short audio segments, ensuring flexibility for future research and alternative feature-extraction approaches.

\subsection{Stage 2: Full-Audio Segment Detection Model}
\label{sec:fullaudio}

Since real music tracks have varying lengths and diverse structures, full-audio analysis is necessary for robust AIGM detection performance. To overcome these limitations, we introduce the Segment Transformer. We preprocessed audio by dividing tracks into manageable segments with beat tracking model. And we trained structural relationships by two different transformer encoder block. This segmented approach allows for more consistent and accurate detection across different track lengths and musical structures, leading to improved full-audio analysis performance.

\noindent\textbf{Music Segment Preparation.} As shown in the stage-2 architecture in Figure~\ref{fig:segment_transformer}, our full-audio detection approach begins with processing full audio tracks into musically meaningful segments. We first detect downbeats in the music using a beat-tracking algorithm, which allows us to identify the beginning of each bar. We then split the audio into 4-bar units, creating a sequence of music segments that preserves the natural rhythmic structure of the composition. These musically informed segments ensure that our model processes the audio in a way that respects its musical organization.

\noindent\textbf{Segment Transformer.} For full-audio analysis, we introduce the Segment Transformer, which processes sequences of music segments to analyze entire compositions comprehensively. As part of our training approach, while the Segment Transformer is compatible with various feature extractors, we use our segment detection models to capitalize on their domain-specific representations. Their outputs serve as rich segment-level representations, allowing the Segment Transformer to learn temporal dependencies and structural coherence across complete musical works.

The Segment Transformer employs a dual-pathway architecture with two parallel transformer encoder structures that capture complementary aspects of musical structure. The first encoder processes content embeddings directly, capturing the semantic and acoustic properties of individual music segments.  The second encoder analyzes global structural patterns with a self-similarity matrix between all segment embeddings. This methods used in many musical structure analysis~\cite{shiu2006similarity}, enabling the model to understand repetitive structures, variations, and overall compositional organization. The outputs from both encoders are then concatenated to form a unified representation that combines local segment characteristics with global structural information. This provides a comprehensive understanding of the entire musical composition for more accurate AIGM detection.

\section{Experiments and Results}

\setlength{\arrayrulewidth}{0.4pt}
\arrayrulecolor{black}
\newcolumntype{?}{!{\color{black}\vrule width 0.8pt}}

\begin{table*}[t]
  \centering
  \caption{Performance comparison of our feature extractor-based models and conventional AI-generated music detection approaches in the Stage 1 pipeline: on FakeMusicCaps (left) using 10-second audio clips, and on SONICS (right) using 5-second segments from each track. Best: \textbf{Bold}; Second best: \underline{Underline}.}
  \renewcommand{\arraystretch}{1.0}
  \setlength{\tabcolsep}{4pt}
  \resizebox{\textwidth}{!}{%
  \begin{tabular}{lcccccc ? lcccccc}
    \toprule
    \multicolumn{7}{c?}{\textbf{FakeMusicCaps Dataset~\cite{fakemusiccaps}}} &
    \multicolumn{7}{c}{\textbf{SONICS Dataset~\cite{rahman2024sonics} (5-sec Segment)}} \\
    \midrule
    \rowcolor{gray!10}
    \multicolumn{7}{l?}{\textit{Baselines}} &
    \multicolumn{7}{l}{\textit{Baselines}} \\
    \textbf{Model} & ACC & Prec. & Recall & F1 & AUC & Spec. 
    & \textbf{Model} & ACC & Prec. & Recall & F1 & AUC & Spec. \\
ResNet18~\cite{he2016deep} & 0.924 & – & – & 0.924 & – & – 
      & ConvNeXt~\cite{liu2022convnet}     & –  & –  & 0.82 & 0.90 & – & 0.98 \\
    SeNet~\cite{hu2018squeeze}    & 0.923 & – & – & 0.923 & – & – 
      & ViT~\cite{dosovitskiy2020image}          & –  & –  & 0.80 & 0.79 & – & 0.79 \\
    MobileNet~\cite{howard2017mobilenets} & 0.968 & – & – & 0.968 & – & – 
      & EfficientViT~\cite{cai2023efficientvit} & –  & –  & 0.78 & 0.87 & – & 0.98 \\
    CNN+LSTM~\cite{hochreiter1997long} & 0.916 & – & – & 0.917 & – & –   & SpecTTTra-$\gamma$~\cite{rahman2024sonics} & – & – & 0.63 & 0.76 & – & 0.98 \\
    Q-SVM~\cite{singh2021detection} & 0.914 & – & – & 0.924 & – & –   & SpecTTTra-$\beta$~\cite{rahman2024sonics} & – & – & 0.69 & 0.78 & – & 0.94 \\
    ViT~\cite{dosovitskiy2020image} & 0.854 & – & – & 0.852 & – & – & SpecTTTra-$\alpha$~\cite{rahman2024sonics} & – & – & 0.71 & 0.80 & – & 0.92 \\
    \midrule
    \rowcolor{lightblue}
    \multicolumn{7}{l?}{\textit{(Ours) Stage 1: Short Audio Segment Detection}} &
    \multicolumn{7}{l}{\textit{(Ours) Stage 1: Short Audio Segment Detection}} \\
    \textbf{Feature Extractor} & ACC & Prec. & Recall & F1 & AUC & Spec. 
    & \textbf{Feature Extractor} & ACC & Prec. & Recall & F1 & AUC & Spec. \\
    Wav2vec 2.0 & \underline{0.980} & \underline{0.988} & \underline{0.988} & \underline{0.988} & \underline{0.944} & \underline{0.939} 
      & Wav2vec 2.0 & 0.995 & 0.991 &\textbf{0.999} & \underline{0.995} & \underline{0.996} & 0.990 \\
    Music2vec    & 0.919 & 0.948 & 0.956 & 0.952 & 0.937 & 0.730 
      & Music2vec    & \underline{0.996} & \underline{0.997} & \underline{0.997} & \textbf{0.997} & \textbf{1.000} & \underline{0.996} \\
    MERT~\cite{mert}          & \textbf{0.994} & \textbf{0.996} & \textbf{0.997} & \textbf{0.996} & \textbf{0.998} & \textbf{0.980} 
      & MERT        & \textbf{0.997} &\textbf{0.998} & 0.996 & \textbf{0.997} &\textbf{1.000} &\textbf{0.998} \\
    FXencoder    & 0.860 & 0.948 & 0.881 & 0.913 & 0.924 & 0.751 
      & FXencoder   & 0.971 & 0.950 & 0.984 & 0.966 &0.993 &0.961\\
    \bottomrule
  \end{tabular}}%
 \label{tab:segment_detection_result}
\end{table*}

\subsection{Datasets} We used the FakeMusicCaps~\cite{fakemusiccaps} and SONICS~\cite{rahman2024sonics} datasets for training and evaluating the stage-1 segment detection model. The FakeMusicCaps dataset consists of 10-second short audio generated by five different Text-to-Music (TTM) models, each employing distinct audio compression techniques and text conditioning mechanisms to produce diverse musical outputs from the same textual descriptions. Specifically, MusicGen~\cite{copet2023simple}, MusicLDM~\cite{chen2024musicldm}, AudioLDM2~\cite{liu2024audioldm}, Stable Audio Open~\cite{evans2025stable}, and Mustango~\cite{melechovsky2023mustango}. The dataset contains a total of 5,373 real and 27,605 AI-generated tracks. 
The SONICS dataset comprises 48,090 real tracks collected dynamically from YouTube using lyrics, titles, and artist metadata sampled from the Genius Song Lyrics dataset, and 49,074 AI-generated tracks, produced with SunoSuno\footref{fn:suno} and Udio\footref{fn:udio} generative models. All tracks have an average duration of 176 seconds. Moreover, for model training, both real and AI-generated tracks were resampled to 16 kHz—consistent with the baseline experiments~\cite{rahman2024sonics}—and the dataset was configured and used in the same way.

\subsection{Implementation Details}
We split the FakeMusicCaps dataset into training, validation, and test sets using an 8:1:1 ratio. For data augmentation, we randomly applied pitch shifting (±2 semitones)
and time stretching (0.8× and 1.25×) with a 50 \% probability was performed exclusively during the training of the SSL-based feature extractor in the segment detection model. Because FXencoder is pre-trained to extract audio features and used with frozen weights, applying augmentations such as pitch shifting or time stretching could distort the original audio patterns. Therefore, we applied data augmentation only to the SSL-based feature extractors, whose weights are updated during training. 
For our segment detection experiments, input audio was resampled to 16 kHz for Wav2vec and Music2vec, 24 kHz for MERT, and 44.1 kHz stereo for the FXencoder. AudioCAT was trained for 30 epochs with a batch size of 8 using binary cross-entropy loss, while the FXencoder was trained for 50 epochs with a batch size of 32 using focal loss. All models were optimized with the Adam optimizer (learning rate 1 × $10^{-5}$, weight decay 1 × $10^{-6}$) and employed early stopping. 

For the Segment Transformer model, we used the SONICS dataset. Each track was divided into a variable number of segments based on its structure, and these segments were then padded or cropped to a fixed sequence length of 48 for input to the transformer model. To extract downbeat information, we used the Beat this!~\cite{foscarin2024beat} model. The downbeat sequence was quantized into an arithmetic sequence aligned with the music's structural rhythm. This model was trained for 50 epochs using binary cross-entropy loss and the Fused Adam optimizer(learning rate of 1 × $10^{-5}$, a weight decay of 1 × $10^{-6}$), and applied early stopping.

All models were trained on a single NVIDIA RTX 5090 GPU. We evaluated their performance on the test set using the following metrics: Accuracy, Precision (Prec.), Recall (Sensitivity), F1-score, area under the ROC curve (AUC), and Specificity (Spec.).

\subsection{Segment Detection Performances}
As shown in Table~\ref{tab:segment_detection_result}, the models proposed in this study outperform the CNN-based MobileNet~\cite{howard2017mobilenets}, which represents the state of the art (SOTA) on the FakeMusicCaps dataset, in every evaluation metric. Our four proposed models consistently outperform the baseline on nearly every metric. FXencoder~\cite{koo2023music}, which emphasizes mixing and mastering characteristics, shows relatively lower performance compared to other models, it still outperforms the ViT~\cite{dosovitskiy2020image} baseline in both accuracy and F1 score. Despite these mixed results, FXencoder demonstrates certain music structural advantages in full-audio detection experiments. On the SONICS dataset, our approaches again demonstrate uniformly superior performance. In particular, MERT~\cite{mert} achieves optimum results among the models, owing to its rich representations through large-scale pre-training on diverse music and audio tasks. 

The improvements are achieved by employing domain-specific encoders—SSL models for general audio understanding and FXencoder for music production details—together with well-matched architectural designs.

\subsection{Full Audio Detection Performances}
For the full audio detection model Segment Transformer, we trained and evaluated our model using SONICS dataset. We used the benchmark reported in~\cite{rahman2024sonics}, specifically for the task of full audio (120-second) AIGM detection in the SONICS dataset.

\begin{table}[t]
  \centering
  \caption{Performance comparison of full-audio segment detection of segment-level detectors on SONICS using 120-second segments from each track. Best: \textbf{Bold}; Second best: \underline{Underline}.}
  \setlength{\tabcolsep}{4pt}
  \setlength{\arrayrulewidth}{0.6pt}
  \resizebox{\columnwidth}{!}{%
  \begin{tabular}{@{}lcccccc@{}}
    \toprule
    \multicolumn{7}{c}{\textbf{SONICS Dataset~\cite{rahman2024sonics} ($120$-sec Segment)}}\\
    \midrule
    \rowcolor{gray!10}
    \multicolumn{7}{l}{\textit{Baselines}}\\
    \textbf{Model} & ACC & Prec. & Recall & F1-Score & AUC & Spec.\\
    ConvNeXt~\cite{liu2022convnet}               & -- & -- & $0.95$ & $0.96$ & -- & $0.98$\\
    ViT~\cite{dosovitskiy2020image}              & -- & -- & $0.82$ & $0.89$ & -- & $0.98$\\
    EfficientViT~\cite{cai2023efficientvit}      & -- & -- & $0.94$ & $0.95$ & -- & $0.97$\\
    SpecTTTra-$\gamma$~\cite{rahman2024sonics}   & -- & -- & $0.79$ & $0.88$ & -- & \underline{$0.99$}\\
    SpecTTTra-$\beta$~\cite{rahman2024sonics}    & -- & -- & $0.86$ & $0.92$ & -- & \underline{$0.99$}\\
    SpecTTTra-$\alpha$~\cite{rahman2024sonics}   & -- & -- & $0.96$ & $0.97$ & -- & \underline{$0.99$}\\
    \midrule
    \rowcolor{lightblue}
    \multicolumn{7}{l}{\textit{(Ours) Stage 2: Segment Transformer for Full Audio Detection}}\\
    \textbf{Feature Extractor} & ACC & Prec. & Recall & F1-Score & AUC & Spec.\\
    Wav2vec 2.0  & $0.7473$ & $0.7127$ & $0.8726$ & $0.7846$ & $0.7412$ & $0.6075$\\
    Music2vec    & $0.9555$ & $0.9609$ & $0.9546$ & $0.9577$ & $0.9907$ & $0.9566$\\
    MERT         & \textbf{$0.9992$} & \textbf{$0.9988$} & \textbf{$0.9996$} & \textbf{$0.9992$} & \textbf{$0.9999$} & \textbf{$0.9987$}\\
    FXencoder    & \underline{$0.9955$} & \underline{$0.9953$} & \underline{$0.9961$} & \underline{$0.9957$} & \underline{$0.9995$} & \underline{$0.9948$}\\
    \bottomrule
  \end{tabular}}%
  \label{tab:model_comparison}
\end{table}

These results can be interpreted through the architectural design of our Segment Transformer, which focuses on modeling temporal connections between musical segments. The performance disparity across different feature extractors reveals the importance of music-specific representations for segment-level modeling. 

Wav2vec 2.0, being primarily designed for speech representations, resulting in significantly degraded performance. In contrast, music-specific feature extractors such as MERT and FXencoder demonstrate exceptional performance, with MERT achieving near-perfect results across all metrics. This suggests that when combined with musically-informed representations, Segment Transformer can effectively learn temporal segment relationships to achieve superior performance.

\section{Conclusions}
In this paper, we presented a novel approach for detecting AI-generated music (AIGM), addressing the needs for music information retrieval domain. Our key contributions are: (1) AudioCAT, a framework that applies cross-attention mechanisms with diverse SSL-based feature extractors and is designed to accommodate any encoder. (2) Full-audio analysis approach using a segment transformer architecture that preserves structural relationships, and a two-stage robust model framework for AIGM detection with strong performance. By considering both complete musical compositions and isolated segments, our approach effectively captures the music structural difference for distinguishing human-composed from AIGM. Experimental results show that our two-stage processing architecture significantly outperforms existing methods. These findings suggest that SSL models designed for the unique characteristics of musical data can serve as an effective means of enhancing detection performance.

Future work could explore alternative methodological approaches, including end-to-end architectures that directly process full-length audio, as well as investigating different fusion strategies for combining segment-level and track-level information. Additionally, exploring other neural architectures beyond transformers may yield further improvements in detection accuracy and computational efficiency.
This research marks a meaningful step toward establishing effective detection methods that can keep pace with the rapid growth of generative AI in music, addressing increasing concerns around copyright and creative authenticity.

\printbibliography

\end{document}